\documentclass[12pt,a4paper]{article}

\usepackage{graphicx}

\begin{document}

\hyphenation{Semprini Neubert Sachrajda}

\noindent {\centering \large \bf \boldmath
        Bayesian analysis of the constraints on $\gamma$ \\
        from $B \rightarrow K\pi$ rates and CP asymmetries \\
        in the flavour-SU(3) approach \\}

\bigskip

\noindent {\centering
M.~Bargiotti,$^1$A.~Bertin,$^1$ M.~Bruschi,$^1$ M.~Capponi,$^1$
                                               S.~De~Castro,$^1$ \\
L.~Fabbri,$^1$ P.~Faccioli,$^1$ D.~Galli,$^1$ B.~Giacobbe,$^1$
                                                 U.~Marconi,$^1$ \\
I.~Massa,$^1$ M.~Piccinini,$^1$ M.~Poli,$^2$ N.~Semprini~Cesari,$^1$
                                                  R.~Spighi,$^1$ \\
V.~Vagnoni,$^1$ S.~Vecchi,$^1$ M.~Villa,$^1$ A.~Vitale,$^1$
                                              and A.~Zoccoli$^1$ \\
}

\smallskip

\noindent {\centering \footnotesize
$^1$\emph{Dipartimento di Fisica dell'Universit\`a di Bologna
                   and INFN Sezione di Bologna - Bologna, Italy} \\
$^2$\emph{Dipartimento di Energetica `Sergio Stecco'
                   dell'Universit\`a di Firenze - Firenze, Italy \\
                   and INFN Sezione di Bologna - Bologna, Italy} \\
}

\bigskip

\begin{abstract}
The relation between the branching ratios and direct CP asymmetries of $B
\rightarrow K\pi$ decays and the angle $\gamma$ of the CKM unitarity
triangle is studied numerically in the general framework of the SU(3)
approach, with minimal assumptions about the parameters not fixed by
flavour-symmetry arguments. Experimental and theoretical uncertainties are
subjected to a statistical treatment according to the Bayesian method. In
this context, the experimental limits recently obtained by CLEO, BaBar and
Belle for the direct CP asymmetries are translated into the bound $|\gamma -
90^{\circ}| > 21^{\circ}$ at the 95\% C.L.. A detailed analysis is carried
out to evaluate the conditions under which measurements of the CP averaged
branching ratios may place a significant constraint on $\gamma$. Predictions
for the ratios of charged ($R_{\mathrm{c}}$) and neutral ($R_{\mathrm{n}}$)
$B \rightarrow K\pi$ decays are also presented.
\end{abstract}

\bigskip

\section{Introduction}

The charmless two-body decays of $B$ mesons play a central role in the
prospect of probing the Standard Model (SM) picture of CP violation. The
well-known $B$-factory benchmark mode $B^{0} \rightarrow \pi^{+}\pi^{-}$
should enable a quite precise measurement of the CKM angle $\alpha$,
provided the unknown penguin contribution to the amplitudes determining the
observable mixing-induced CP asymmetry can be controlled with sufficient
accuracy through additional measurements of the three isospin-related $B
\rightarrow \pi\pi$ branching ratios \cite{GroLon90} (including the very
challenging $B^{0} \rightarrow \pi^{0}\pi^{0}$ mode). The SU(3) flavour-
symmetry relation between $B^{0} \rightarrow \pi^{+}\pi^{-}$ and $B^{0}
\rightarrow K^{0}\overline{K^{0}}$ offers an alternative way of reducing
penguin uncertainties in the extraction of $\alpha$ \cite{BurFl95}.
Furthermore, a time dependent analysis combining $B^{0} \rightarrow
\pi^{+}\pi^{-}$ with its `U-spin' counterpart ($d \leftrightarrow s$) $B_{s}
\rightarrow K^{+}K^{-}$ can determine the other two CKM angles $\beta$ and
$\gamma$ simultaneously \cite{Fleischer99} (there is moreover a variant of
this method, replacing $B_{s} \rightarrow K^{+}K^{-}$ with $B^{0}
\rightarrow K^{\pm}\pi^{\mp}$ \cite{Fleischer00}). Further information on
$\gamma$ and/or $\beta$ can be obtained from the direct and mixing-induced
asymmetries in $B^{0} \rightarrow K^{0}_{S}\pi^{0}$
\cite{AliKramerLu98,BurFl98,BurFl00}.

\begin{table}[tb]
\caption{Measurements of the $B \rightarrow K\pi$ branching ratios
($\mathcal{B}$) and direct CP asymmetries ($a_{\mathrm{CP}}$). The averages
have been computed combining statistical and systematic errors in quadrature
and taking into account asymmetric errors; correlations between the
systematical errors have been neglected. The list includes the measurement
of the $\pi^{\pm} \pi^{0}$ branching ratio, used as an ingredient of the
SU(3) analysis.}
\label{data}
\renewcommand{\arraystretch}{1.3}
\centering
\begin{tabular}{l|ccc|c}
\hline\hline
$\mathcal{B} \times 10^{-6}$ & CLEO \cite{CLEO} & Belle \cite{Belle} &
BaBar \cite{BaBar} & \ \ Average\ \  \\

\hline
$K^{0} \pi^{\pm}$ & $18.2\ ^{+4.6}_{-4.0}\pm1.6$ & $13.7\
^{+5.7}_{-4.8}\ ^{+1.9}_{-1.8}$ & $18.2\ ^{+3.3}_{-3.0}\pm2.0$ & $17.4
\pm 2.6$ \\

$K^{\pm} \pi^{0}$ & $11.6\ ^{+3.0}_{-2.7}\ ^{+1.4}_{-1.3}$ & $16.3\
^{+3.5}_{-3.3}\ ^{+1.6}_{-1.8}$ & $10.8\ ^{+2.1}_{-1.9} \pm 1.0$ &
$12.1 \pm 1.7$ \\

$K^{0} \pi^{0}$ & $14.6\ ^{+5.9}_{-5.1}\ ^{+2.4}_{-3.3}$ & $16.0\
^{+7.2}_{-5.9}\ ^{+2.5}_{-2.7}$ & $8.2\ ^{+3.1}_{-2.7} \pm 1.2$ &
$10.8 \pm 2.7$ \\

$K^{\pm} \pi^{\mp}$ & $17.2\ ^{+2.5}_{-2.4}\pm1.2$ & $19.3\
^{+3.4}_{-3.2}\ ^{+1.5}_{-0.6}$ & $16.7 \pm 1.6 \pm 1.3$ & $17.4 \pm
1.5$ \\

$\pi^{\pm} \pi^{0}$ & $5.6\ ^{+2.6}_{-2.3}\pm1.7$ & $7.8\
^{+3.8}_{-3.2}\ ^{+0.8}_{-1.2}$ & $5.1\ ^{+2.0}_{-1.8}\pm0.8$ & $5.8
\pm 1.5$ \\

\hline

$a_{\mathrm{CP}} \times 10^{-2}$ & & & & \\

\hline

$K^{0} \pi^{\pm}$   & $+18 \pm 24$ & $+9.8\ ^{+43.0}_{-34.3}\
^{+2.0}_{-6.3}$ & $-21 \pm 18 \pm 3$ & $-4 \pm 13$ \\

$K^{\pm} \pi^{0}$   & $-29 \pm 23$ & $-5.9\ ^{+22.2}_{-19.6}\
^{+5.5}_{-1.7}$ & $0 \pm 18 \pm 4$ & $-10 \pm 12$ \\

$K^{\pm} \pi^{\mp}$ &  $-4 \pm 16$ & $+4.4\ ^{+18.6}_{-16.7}\
^{+1.8}_{-2.1}$ & $-7 \pm 8 \pm 2$ & $-5 \pm 7$ \\

\hline\hline
\end{tabular}
\end{table}

Much interest has been excited by the possibility of constraining
the CKM phase $\gamma$ using only CP averaged measurements of $B
\rightarrow K\pi$ and $B \rightarrow \pi\pi$ decays
\cite{GroRosLon94, Fleischer95, FlMann97, BurFlMann97,
Fleischer98, NeuRos98, BurFl98, Neub98, BurFl00, Atwood98, He00,
gamgt90deg, BBNS00, KeumLiSanda00, Ciuchini01}. The first
branching ratio measurements by CLEO \cite{CLEO}, now confirmed by
consistent evaluations by Belle \cite{Belle} and BaBar
\cite{BaBar} (see Table~\ref{data}), were often interpreted, in
different theoretical approaches, as an indication that the angle
$\gamma$ may lie in the second quadrant \cite{gamgt90deg}, in
conflict with the expectation $\gamma \simeq 60^\circ$ derived
from global analyses of the unitarity triangle (UT) \cite{CKM}.
Recently, the first results of the experimental search for CP
violation in these decays have appeared, in the form of upper
limits for the direct CP asymmetries (Table~\ref{data});
measurements of these observables will place a straightforward
constraint on the value of the CP violating phase $\gamma$.

In this paper, the implications of the available experimental results on $B
\rightarrow K \pi$ decays are studied in the framework of the flavour-SU(3)
approach \cite{GroRosLon94, Fleischer95, FlMann97, BurFlMann97, Fleischer98,
NeuRos98, BurFl98, Neub98, BurFl00}, where the amount of theoretical input
about QCD dynamics is minimized through the use of flavour symmetry
relations involving additional measured quantities. In particular, the
present analysis will focus on the theoretically clean strategy employing
the ratios of charged and neutral $B$ decay rates defined as \cite{BurFl98}
\begin{eqnarray}
R_{\mathrm{c}} & = & 2 \cdot \frac{\mathcal{B}(B^{+} \rightarrow K^{+}
\pi^{0}) + \mathcal{B}(B^{-} \rightarrow K^{-}
\pi^{0})}{\mathcal{B}(B^{+} \rightarrow K^{0} \pi^{+}) +
\mathcal{B}(B^{-} \rightarrow \overline{K^{0}} \pi^{-})} =
1.34^{+0.30}_{-0.25},  \label{eq:Rc} \\*
R_{\mathrm{n}} & = & \frac{1}{2} \cdot \frac{\mathcal{B}(B^{0}
\rightarrow K^{+} \pi^{-}) + \mathcal{B}(\overline{B^{0}} \rightarrow
K^{-} \pi^{+})} {\mathcal{B}(B^{0} \rightarrow K^{0} \pi^{0}) +
\mathcal{B}(\overline{B^{0}} \rightarrow \overline{K^{0}} \pi^{0})} =
0.73^{+0.24}_{-0.17}
\label{eq:Rn}
\end{eqnarray}
(the experimental averages\footnote{Possible correlations are
neglected; the asymmetric errors and the shift of the central
value with respect to the ratio of the central values derive from
taking into account the effect of a large uncertainty in the
denominator.} obtained from the last column of Table~\ref{data}
are indicated), as well as the direct CP asymmetries
($a_{\mathrm{CP}}$) of the four decay modes, here considered with
the following sign convention:
\begin{eqnarray}
a_{\mathrm{CP}} & = & \frac{\mathcal{B}(\overline{B} \rightarrow
\overline{f}) - \mathcal{B}(B \rightarrow f)}{\mathcal{B}(\overline{B}
\rightarrow \overline{f}) + \mathcal{B}(B \rightarrow f)}. \label{aCP}
\end{eqnarray}
The ratio $[\mathcal{B}(B^{0} \rightarrow K^{+} \pi^{-}) +
\mathcal{B}(\overline{B^{0}} \rightarrow K^{-} \pi^{+})] /
[\mathcal{B}(B^{+} \rightarrow K^{0} \pi^{+}) + \mathcal{B}(B^{-}
\rightarrow \overline{K^{0}} \pi^{-})]$ was also shown to provide a
potentially effective bound on $\gamma$ \cite{FlMann97}. However, the
uncertainty as to whether the contribution of the colour suppressed
electroweak penguin amplitude is negligible and a greater sensitivity to
rescattering effects (see Refs.~\cite{Fleischer98,BurFl98}) make this
`mixed' strategy less model independent. Recently, different approaches have
been developed to evaluate the $B \rightarrow K \pi$ decay amplitudes
through a deeper insight into the details of QCD dynamics \cite{BBNS00,
KeumLiSanda00}. Although these methods may in principle enable improved
bounds to be placed on $\gamma$, they currently rely on theoretical
assumptions which are still a matter of debate \cite{Ciuchini01,
CKMworkshop}.

Limitations to the theoretical accuracy of the SU(3) method adopted in the
present analysis for deriving bounds on $\gamma$ from $R_{\mathrm{c}}$,
$R_{\mathrm{n}}$ and from the direct CP asymmetries would only be
represented by large non-factorizable SU(3)-breaking effects altering the
relative weight of tree and electroweak penguin amplitudes with respect to
the dominant QCD penguin contribution. Parameters not fixed by flavour
symmetry arguments, such as the strong phase differences between tree and
QCD penguin amplitudes and quantities expressing the unknown contribution of
certain rescattering effects, will be treated as free variables of the
numerical analysis.

The aim of the analysis reported in this paper is to estimate whether (or
under what conditions) the comparison between measured $B \rightarrow K \pi$
rates and asymmetries and the corresponding flavour-SU(3) expectations may
serve as a test of the SM. Constraints on $\gamma$ and on the strong phases
are analysed in the framework of Bayesian statistics, where a definite
statistical meaning is assigned both to the experimental data and to the
theoretical ranges which constitute the \emph{a priori} knowledge of the
input parameters. Consequently, the resulting predictions are the expression
of all the experimental and theoretical information assumed as input to the
analysis, differently from previous studies which only considered a number
of illustrative cases corresponding to fixed values of theoretical and
experimental parameters.

The implications of the available measurements and the foreseeable impact of
precise data on the determination of $\gamma$ and of the strong phases are
studied in Sec. \ref{gamma}. Predictions for $R_{\mathrm{c}}$ and
$R_{\mathrm{n}}$ derived by fixing $\gamma$ to the SM expectation are
reported in Sec.~\ref{ratios}, where the sensitivity of the values obtained
to the main experimental and theoretical inputs is evaluated.

\section{\boldmath Constraints on $\gamma$ from $B \rightarrow K\pi$}
\label{gamma}

\subsection{\boldmath Parametrization of the $B \rightarrow K\pi$
decay amplitudes}

Two alternative parametrizations of the $B \rightarrow K\pi$ decay
amplitudes can be found in the literature \cite{BurFl98,Neub98}. The
following analysis assumes the notation used in Ref.~\cite{BurFl98}:
\begin{eqnarray}
\mathcal{A}(B^{+} \rightarrow K^{0} \pi^{+}) & = &
\tilde{P}_{\mathrm{c}} \left[ 1 + \rho_{\mathrm{c}}
e^{i \theta_{\mathrm{c}}} e^{i \gamma} \right], \label{eq:param1}\\*
-\sqrt{2}\mathcal{A}(B^{+} \rightarrow K^{+} \pi^{0}) &
= & \tilde{P}_{\mathrm{c}} \left[ 1 + \rho_{\mathrm{c}}
e^{i \theta_{\mathrm{c}}} e^{i \gamma} + \right. \nonumber \\*
& & - \left. r_{\mathrm{c}} e^{i \delta_{\mathrm{c}}} (e^{i \gamma}-
q e^{i \omega}) \sqrt{1 + 2 \rho_{\mathrm{c}} \cos\theta_{\mathrm{c}}
\cos{\gamma} + \rho_{\mathrm{c}}^{2}} \ \right], \label{eq:param2} \\*
\sqrt{2}\mathcal{A}(B^{0} \rightarrow K^{0} \pi^{0}) & = &
\tilde{P}_{\mathrm{n}} \left[ 1 + \rho_{\mathrm{n}}
e^{i \theta_{\mathrm{n}}} e^{i \gamma} \right], \label{eq:param3}\\*
-\mathcal{A}(B^{0} \rightarrow K^{+} \pi^{-}) & = &
\tilde{P}_{\mathrm{n}} \left[ 1 + \rho_{\mathrm{n}}
e^{i \theta_{\mathrm{n}}} e^{i \gamma} + \right. \nonumber \\*
& & - \left. r_{\mathrm{n}} e^{i \delta_{\mathrm{n}}} (e^{i \gamma}-
q e^{i \omega}) \sqrt{1 + 2 \rho_{\mathrm{n}} \cos\theta_{\mathrm{n}}
\cos{\gamma} + \rho_{\mathrm{n}}^{2}} \ \right], \label{eq:param4}
\end{eqnarray}
where

\begin{list}{-}{}

\item $\tilde{P}_{\mathrm{c}}$, $\tilde{P}_{\mathrm{n}}$ are CP invariant
factors containing the dominant QCD penguin contributions; they cancel out
in the expressions of the ratios $R_{\mathrm{c}}$,
$R_{\mathrm{n}}$ and of the CP asymmetries.

\item the terms $\rho_{\mathrm{c}} e^{i \theta_{\mathrm{c}}} e^{i \gamma}$
and $\rho_{\mathrm{n}} e^{i \theta_{\mathrm{n}}} e^{i \gamma}$
($\theta_{\mathrm{c}}$ and $\theta_{\mathrm{n}}$ are strong phases)
determine the magnitude of the direct CP violation in the decays $B^{+}
\rightarrow K^{0} \pi^{+}$ and $B^{0} \rightarrow K^{0} \pi^{0}$
respectively; they are generally expected to be quite small
[$\rho_{\mathrm{(c,n)}} = \mathcal{O}(10^{-2})$], but their importance may
be enhanced by final state interactions \cite{Fleischer98}. Calculations
performed in the framework of the `QCD factorization' approach \cite{BBNS00}
point to no significant modification due to final state interactions. The
present experimental average $a_{\mathrm{CP}}(K^{0}\pi^{+}) = -0.04 \pm
0.13$ (Table~\ref{data}) is also not in favour of the presence of large
rescattering effects. An upper bound on $\rho_{\mathrm{c}}$ can be deduced
from the experimental limit on the ratio $\mathcal{B}(B^{\pm} \rightarrow
K^{0} K^{\pm})/\mathcal{B}(B^{\pm} \rightarrow K^{0} \pi^{\pm})$
\cite{BurFlMann97,FKNP98} by exploiting the U-spin relation
\begin{equation}
\frac{\mathcal{B}(B^{+} \rightarrow \overline{K^{0}}
K^{+})}{\mathcal{B}(B^{+} \rightarrow K^{0} \pi^{+})} \times
R_{\mathrm{SU}(3)}^2 \bar{\lambda}^2 = \frac{ \rho_{\mathrm{c}}^2 -2
\bar{\lambda}^2 \rho_{\mathrm{c}} \cos \theta_{\mathrm{c}} \cos \gamma
+\bar{\lambda}^4}{1 + 2 \rho_{\mathrm{c}} \cos \theta_{\mathrm{c}}
\cos \gamma + \rho_{\mathrm{c}}^2}, \label{rhoKK}
\end{equation}
where the factor $R_{\mathrm{SU}(3)}$ represents the correction for SU(3)
symmetry breaking, which is of the order of $0.7$ at the factorizable level
\cite{BurFlMann97}, and $\bar{\lambda} \equiv \lambda/(1-\lambda^2/2)$. The
most stringent limit available for the $\overline{K^{0}} K^{+}$ branching
ratio, $\mathcal{B}(B^{+} \rightarrow \overline{K^{0}} K^{+}) < 2.4 \times
10^{-6}$ at the $90\%$ C.L., has been achieved by the BaBar Collaboration
\cite{BaBar}. They also report a fitted value of $(-1.3 ^{+1.4}_{-1.0} \pm
0.7) \times 10^{-6}$; using this value together with the world average for
$\mathcal{B}(B^{+} \rightarrow K^{0} \pi^{+})$ given in Table~\ref{data}, a
fit of Eq. (\ref{rhoKK}), with $R_{\mathrm{SU}(3)} = 0.7$, $\lambda = 0.2224
\pm 0.0020$ \cite{CKM} and flat prior distributions assumed for $\gamma$ and
$\theta_{\mathrm{c}}$ over $[-180^{\circ},180^{\circ}]$, gives the result
\begin{equation}
\rho_{\mathrm{c}} < 0.09 \mathrm{\ \ \ at \ the \ } 95\% \mathrm{\ C.L.}.
\end{equation}
Both $\rho_{\mathrm{c}}$ and $\rho_{\mathrm{n}}$ will be hereafter assumed
to be included in the range $[0,0.2]$, while the strong phases
$\theta_{\mathrm{c}}$ and $\theta_{\mathrm{n}}$ will be treated as free
parameters. To estimate the importance of the resulting uncertainty,
examples assuming $\rho_{\mathrm{(c,n)}} = 0$ will also be considered.

\item $r_{\mathrm{(c,n)}} e^{i \delta_{\mathrm{(c,n)}}}$ and $q e^{i
\omega}$ ($\delta_{\mathrm{(c,n)}}$ and $\omega$ are strong phase
differences) represent, in a simplified description, ratios of
tree-to-QCD-penguin and of electroweak-penguin-to-tree amplitudes,
respectively. In the limit of SU(3) invariance they are estimated
as \cite{GroRosLon94, Fleischer95, NeuRos98, BurFl98}
\begin{eqnarray}
r_{\mathrm{c}} & = & |V_{us}/V_{ud}| \cdot(f_{K}/f_{\pi}) \cdot\sqrt{2
\mathcal{B}(\pi^{\pm} \pi^{0})/\mathcal{B}(K^{0} \pi^{\pm})} \nonumber \\
& = & 0.23 \pm 0.03, \label{rpm} \\
r_{\mathrm{n}} & = & |V_{us}/V_{ud}| \cdot(f_{K}/f_{\pi})
\cdot\sqrt{\mathcal{B}(\pi^{\pm} \pi^{0})/\mathcal{B}(K^{0} \pi^{0})}
\nonumber \\
& = & 0.20 \pm 0.04, \label{r0} \\
q e^{i \omega} & = & \frac{0.057}{|V_{ub}/V_{cb}|} = 0.65 \pm 0.15,
\label{eq:qEW}
\end{eqnarray}
with factorizable SU(3)-breaking corrections included.

\end{list}

Using Eqs. (\ref{eq:param1}, \ref{eq:param2}, \ref{eq:param3},
\ref{eq:param4}), the ratios $R_{\mathrm{c}}$ and $R_{\mathrm{n}}$
[Eqs.~(\ref{eq:Rc}, \ref{eq:Rn})] of CP averaged branching fractions and the
direct CP asymmetries in the four modes [Eq.~(\ref{aCP})], can
therefore be calculated as functions of the parameters
\begin{eqnarray}
r_{\mathrm{c}} e^{i \delta_{\mathrm{c}}} & = & (0.23 \pm 0.03)\ e^{i
[-180,180]^{\circ}}, \nonumber \\
r_{\mathrm{n}} e^{i \delta_{\mathrm{n}}} & = & (0.20 \pm 0.04)\ e^{i
[-180,180]^{\circ}}, \nonumber \\
\rho_{\mathrm{c}} e^{i \theta_{\mathrm{c}}}, \rho_{\mathrm{n}} e^{i
\theta_{\mathrm{n}}} & = & [0,0.2]\ e^{i [-180,180]^{\circ}}, \nonumber \\
q e^{i \omega} & = & 0.65 \pm 0.15 \label{parameters}
\end{eqnarray}
and of the angle $\gamma$. The square brackets indicate that flat
`prior' distributions within the given range are assumed in the
following numerical analysis; the remaining parameters are treated
as Gaussian variables. Constraints on $\gamma$ and on the unknown
CP conserving phases are obtained by fitting the resulting
expressions to the measured branching ratios and CP
asymmetries.\footnote{In view of the correlation of
$r_{\mathrm{c}}$ and $r_{\mathrm{n}}$ between them and with the
values of $\mathcal{B}(K^{0} \pi^{\pm})$ and $\mathcal{B}(K^{0}
\pi^{0})$ [Eqs.~(\ref{rpm}, \ref{r0})], the actual fit procedure
makes use of the four branching ratio measurements instead of the
ratios $R_{\mathrm{c}}$ and $R_{\mathrm{n}}$. This choice also
avoids the problem of dealing with the non-Gaussian distribution
of the measured $R_{\mathrm{c}}$ and $R_{\mathrm{n}}$
[Eqs.~(\ref{eq:Rc}, \ref{eq:Rn})].} The method used in the present
analysis, based on the Bayesian inference model, is the same
described and discussed in Refs.~\cite{CKM, CKMworkshop} in
connection with its application to the CKM fits.

\subsection{Constraints from branching ratio measurements}

\begin{figure}[htb]
\centering
\includegraphics[width=0.6\textwidth]{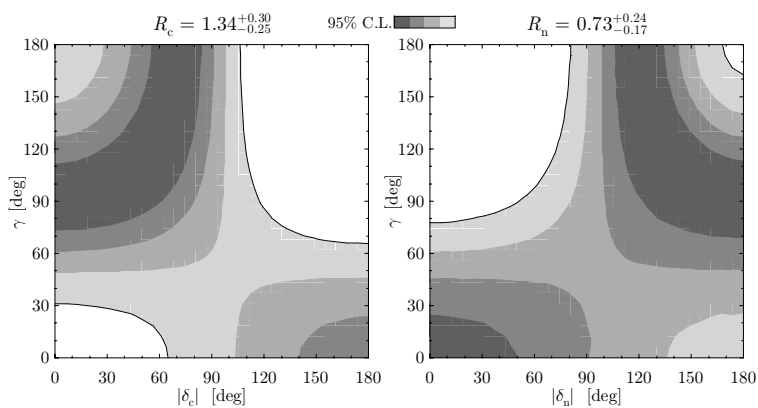}
\caption{Constraints determined on the $|\delta_{\mathrm{(c,n)}}|,\gamma$
plane by the present CP averaged data on $B \rightarrow K \pi$.}
\label{fig:gammadeltaR_today}
\end{figure}

The constraints determined by the current measurements of
$R_{\mathrm{c}}$ and $R_{\mathrm{n}}$ on $\gamma$ and on the
strong phases $\delta_{\mathrm{c}}$ and $\delta_{\mathrm{n}}$ are
represented in Fig.~\ref{fig:gammadeltaR_today}. The shaded areas
plotted in the $|\delta_{\mathrm{c}}|,\gamma$ and
$|\delta_{\mathrm{n}}|,\gamma$ planes\footnote{The constraints are
symmetrical with respect to both $\gamma = 0$ and
$\delta_{\mathrm{(c,n)}} = 0$. The angle $\gamma$ is anyway
assumed to belong to the first two quadrants, as implied by the
positive sign of the $K^{0} - \overline{K^{0}}$ mixing parameter
$B_K$.} are regions of 95\% probability, where darker shades
indicate higher values of the p.d.f. for the two variables. As can
be seen, the current data are not precise enough to place a bound
on $\gamma$ independently of the value of the strong phases (or
vice versa). For comparison, the SM expectation for $\gamma$
resulting from global fits of UT constraints \cite{CKM} is
included between $40^{\circ}$ and $80^{\circ}$, with slight
differences in the exact range depending on the choice of inputs
and on the statistical method used. For the purposes of the
present analysis, the determination
\begin{equation}
\gamma_{_{\mathrm{CKM}fit}} = (56 \pm 8)^{\circ} \label{eq:gamma}
\end{equation}
will be assumed.

\begin{figure}[p]
\centering
\includegraphics[width=\textwidth]{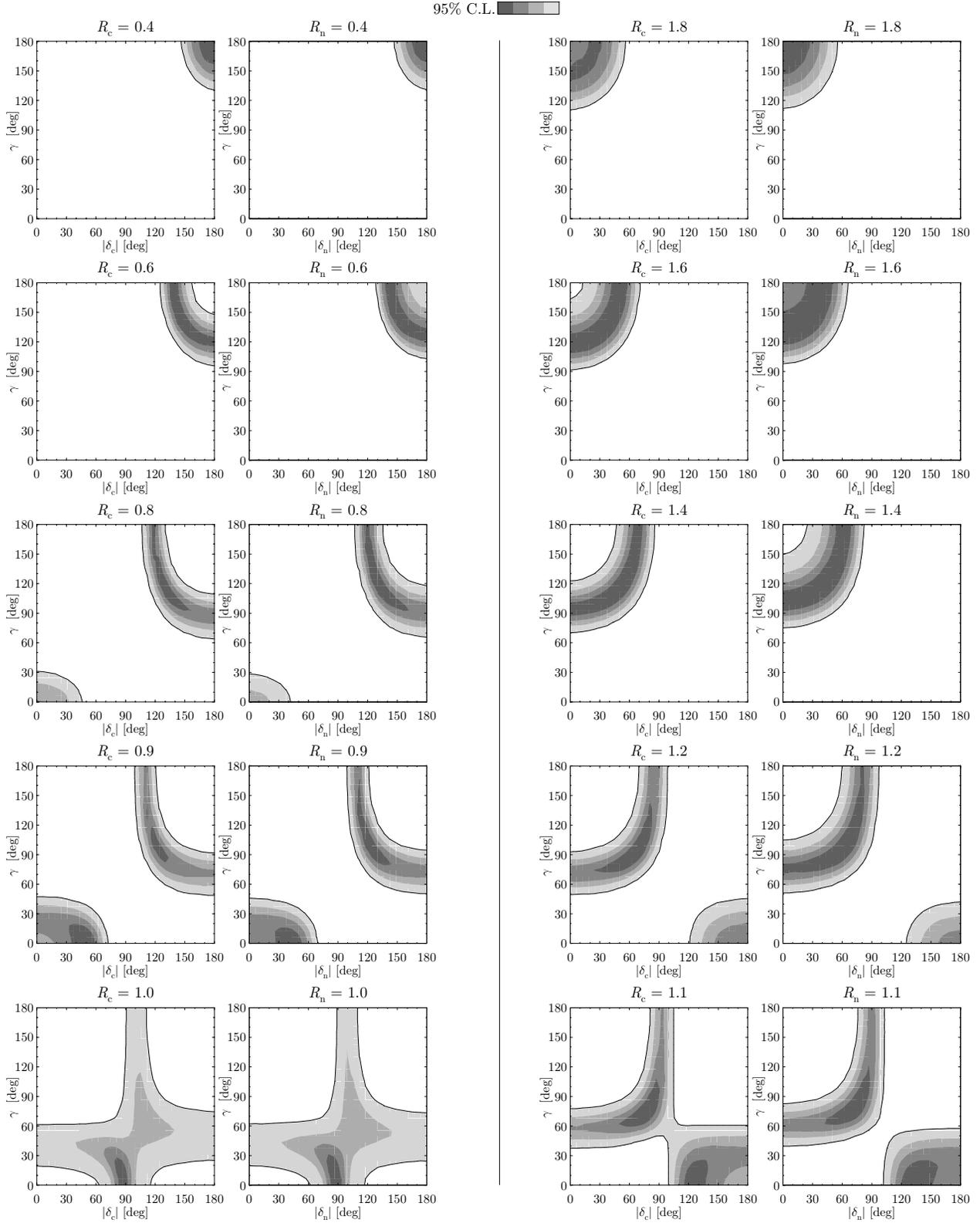}
\caption{Constraints on the $|\delta_{\mathrm{(c,n)}}|,\gamma$ plane
obtainable from precise measurements of $R_{\mathrm{n}}$ and
$R_{\mathrm{c}}$.}
\label{fig:gammadeltaR_max}
\end{figure}

\begin{figure}[htb]
\centering
\includegraphics[width=0.6\textwidth]{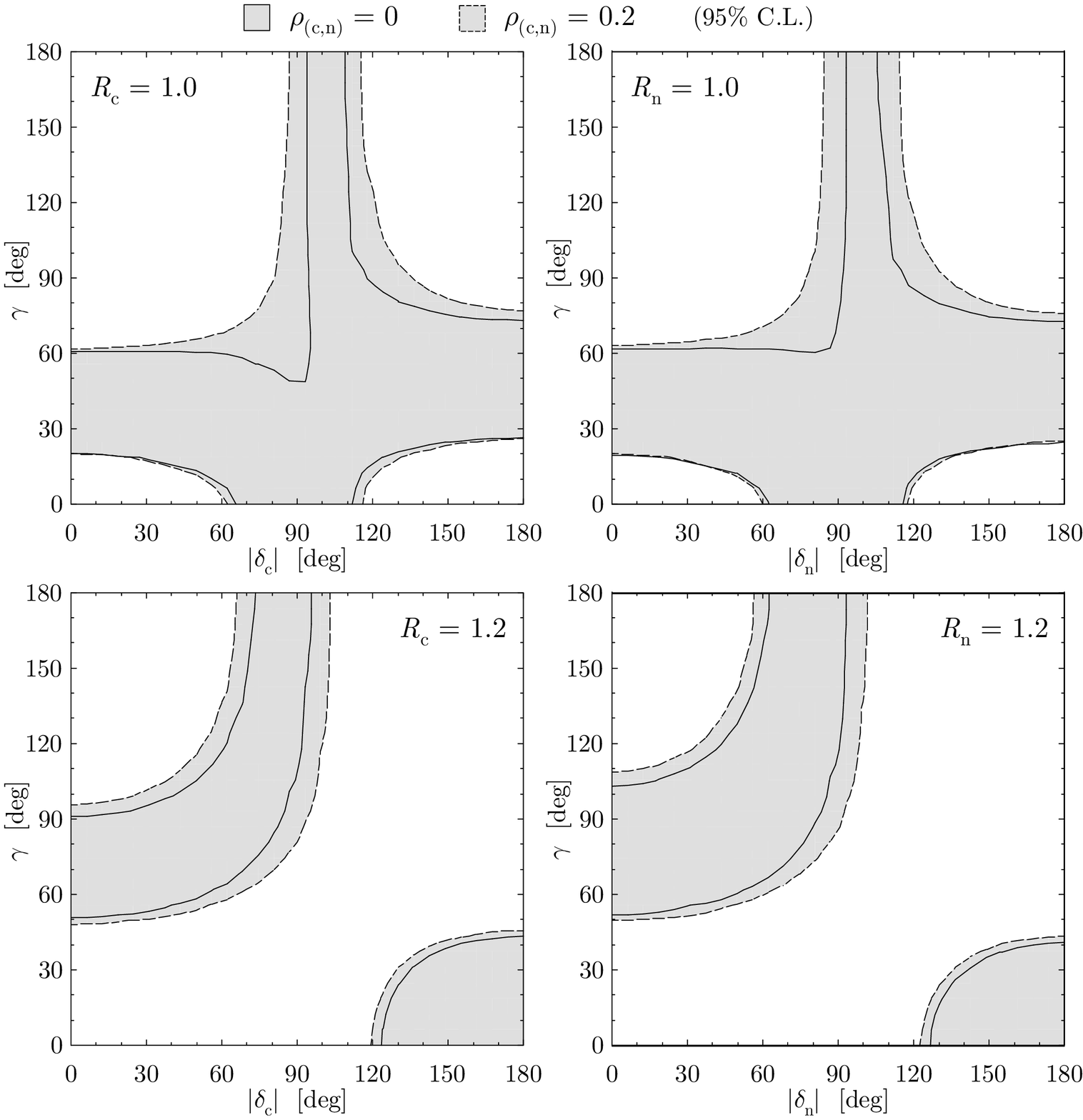}
\caption{Examples of constraints on $\gamma$ and
$\delta_{\mathrm{(c,n)}}$ obtainable from precise measurements of
$R_{\mathrm{(c,n)}} = 1.0$ and $R_{\mathrm{(c,n)}} = 1.2$: the
results in the absence of appreciable rescattering effects
($\rho_{\mathrm{(c,n)}} \simeq 0$) are compared with those
obtained with the maximum value of $\rho_{\mathrm{(c,n)}}$ in the
range assumed in the present analysis ($\rho_{\mathrm{(c,n)}} =
0.2$). The contours, solid and dashed respectively, delimit
regions allowed at the 95\% C.L. in the two scenarios.}
\label{fig:gammadeltaR_rho0}
\end{figure}

Figure~\ref{fig:gammadeltaR_max} shows examples of how the determination of
$\gamma$ and of the strong phases can evolve with improved measurements of
the CP averaged branching ratios. For each value of $R_{\mathrm{c}}$ and
$R_{\mathrm{n}}$, supposed to be measured with negligible experimental
uncertainty,\footnote{An improvement of the present precision by about one
order of magnitude would fulfil such condition.} the graphs represent
regions allowed at the $95\%$ C.L. in the $|\delta_{\mathrm{(c,n)}}|,\gamma$
plane, as determined exclusively by the indeterminacy assumed for the
remaining parameters $q e^{i \omega}$, $r_{\mathrm{(c,n)}}$ and
$\rho_{\mathrm{(c,n)}} e^{i \theta_{\mathrm{(c,n)}}}$
[Eq.~(\ref{parameters})]. The following prospects can be outlined:

\begin{itemize}

\item  A measured value of $R_{\mathrm{(c,n)}}$ smaller than $0.8$
determines a lower bound on both $\gamma$ and
$|\delta_{\mathrm{(c,n)}}|$; the excluded ranges increase with
decreasing value of $R_{\mathrm{(c,n)}}$. In an almost symmetrical
way, if $R_{\mathrm{(c,n)}}$ assumes a value greater than $1.2$, it is
possible to put a lower bound on $\gamma$ and an \emph{upper} bound on
$|\delta_{\mathrm{(c,n)}}|$. On the other hand, a precise measurement
of $R_{\mathrm{(c,n)}}$ having a value included between 0.8 and 1.2
would not be sufficient by itself to delimit ranges of preferred values
for $\gamma$ and $|\delta_{\mathrm{(c,n)}}|$.

\item At the same time, measurements with values diverging from
$R_{\mathrm{(c,n)}} \simeq 1$ are progressively in favour of the range
$\gamma > 90^{\circ}$ and therefore in contrast with the current SM
determination of the UT [Eq.~(\ref{eq:gamma})].

\item By fixing the strong phases (they can actually be calculated
using different theoretical techniques
\cite{AliKramerLu98,BBNS00,KeumLiSanda00}), it becomes possible to
determine $\gamma$ even in the least favourable case of
measurements of $R_{\mathrm{c}}$ and $R_{\mathrm{n}}$ with values
near 1. To illustrate with an example, a sensitivity of $\Delta
\gamma \sim 10^{\circ}$ can be reached, in that peculiar case, by
constraining the strong phases $\delta_{\mathrm{c}}$ and
$\delta_{\mathrm{n}}$ into the (hypothetical) range
$[-30^{\circ},30^{\circ}]$.

\item One interesting prospect is connected with the determination of the
strong phases. It was pointed out \cite{BurFl00} that the first measurements
of $R_{\mathrm{c}}$ and $R_{\mathrm{n}}$ by CLEO favoured values of
$\delta_{\mathrm{c}}$ and $\delta_{\mathrm{n}}$ which were markedly
different from each other, in conflict with the approximate expectation
$\delta_{\mathrm{c}} \simeq \delta_{\mathrm{n}}$. As can be seen in
Fig.~\ref{fig:gammadeltaR_today}, no discrepancy between the values of
$\delta_{\mathrm{c}}$ and $\delta_{\mathrm{n}}$ is implied by the present
data. However, improved measurements will have the potential to establish
such a contradiction: for example, measurements of $R_{\mathrm{c}}$ and
$R_{\mathrm{n}}$ confirming the present central values, respectively greater
than $1.2$ and smaller than $0.8$, at a sufficient level of precision (see
Fig.~\ref{fig:gammadeltaR_max}) would point definitely to
$|\delta_{\mathrm{c}}| < 90^{\circ}$ and $|\delta_{\mathrm{n}}| > 90^{\circ}$.

\end{itemize}

In the above examples, the unknown contribution of final state interactions
to the $K^{0} \pi^{+}$ and $K^{0} \pi^{0}$ decay amplitudes has been
parametrized by allowing $\rho_{\mathrm{c}}$ and $\rho_{\mathrm{n}}$ to be
as large as $0.2$. A comparison between the most favourable scenario
assuming $\rho_{\mathrm{c}} = \rho_{\mathrm{c}} = 0$ and the one in which
$\rho_{\mathrm{c}}$ and $\rho_{\mathrm{c}}$ are fixed to $0.2$ is shown in
Fig.~\ref{fig:gammadeltaR_rho0} for the two representative cases
$R_{\mathrm{(c,n)}} = 1.0$ and $R_{\mathrm{(c,n)}} = 1.2$. As can be seen,
the shape of the constraints remains almost the same in the two cases.
Apparently, no crucial improvement in the determination of $\gamma$ at a
fixed value of $\delta_{\mathrm{(c,n)}}$ could be obtained by further
reducing the uncertainties related to the magnitude of the rescattering
effects.

\subsection{Constraints from CP asymmetries}

\begin{figure}[htb]
\centering
\includegraphics[width=0.55\textwidth]{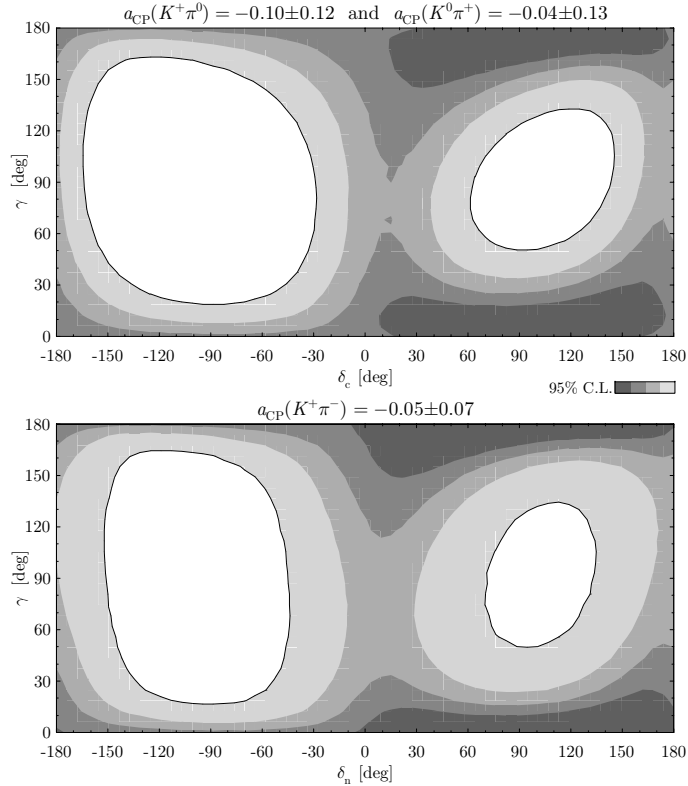}
\caption{Constraints determined by the current CP asymmetry measurements:
allowed regions in the $\delta_{\mathrm{(c,n)}},\gamma$ plane.}
\label{fig:gammadeltaA_today}
\end{figure}

\begin{figure}[htb]
\centering
\includegraphics[width=0.55\textwidth]{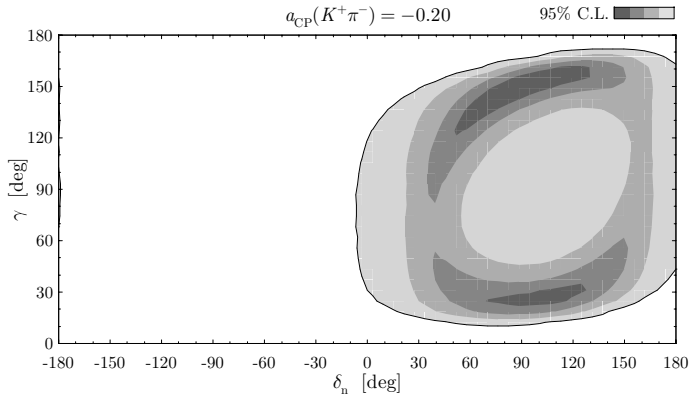}
\caption{Example of constraint on the $\delta_{\mathrm{n}},\gamma$
plane as determined by a precise measurement of the direct CP
asymmetry $a_{\mathrm{CP}}(K^+\pi^-)$.}
\label{fig:gammadeltaA-020}
\end{figure}

\begin{figure}[htb]
\centering
\includegraphics[width=0.40\textwidth]{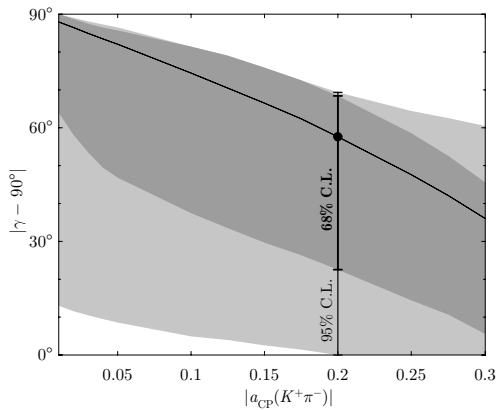}
\caption{Ranges allowed for $|\gamma-90^{\circ}|$ at the 68\% and 95\% C.L.
as a function of the measured value of $a_{\mathrm{CP}}(K^+\pi^-)$.}
\label{fig:gammaA_max}
\end{figure}

\begin{figure}[htb]
\centering
\includegraphics[width=0.55\textwidth]{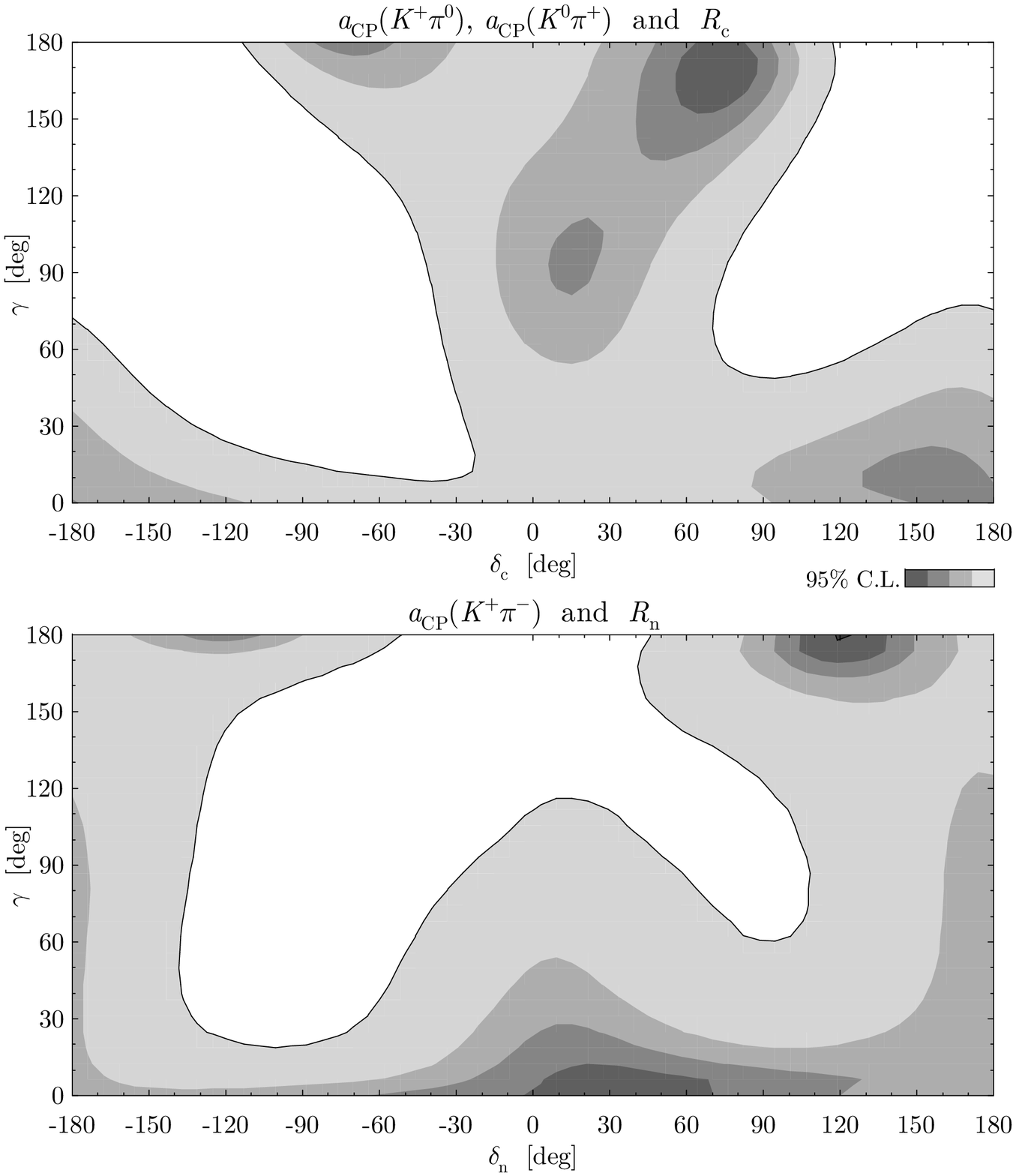}
\caption{Preferred regions in the $\delta_{\mathrm{c}},\gamma$ and
$\delta_{\mathrm{n}},\gamma$ planes, as determined by a global fit of the
present $B \rightarrow K\pi$ rate and CP asymmetry measurements.}
\label{fig:gammadelta_R+A_today}
\end{figure}

The constraints provided in the $\delta_{\mathrm{(c,n)}},\gamma$ plane by
the first experimental results on the direct CP asymmetries are shown in
Fig.~\ref{fig:gammadeltaA_today}. Here $a_{\mathrm{CP}}(K^+\pi^0)$ and
$a_{\mathrm{CP}}(K^0\pi^+)$, both depending on the same subset of parameters
(the one including $\rho_{\mathrm{c}} e^{i \theta_{\mathrm{c}}}$), are
represented as a single constraint. The present data, still consistent with
zero CP violation, determine an upper bound on $\gamma$ in the first
quadrant and a lower bound in the second quadrant, almost symmetrically with
respect to $\gamma=90^{\circ}$. The limit
\begin{equation}
|\gamma-90^{\circ}|>21^{\circ} \mathrm{\ \ at\ the\ } 95\% \mathrm{\ C.L.}
\end{equation}
can be derived from a fit of both constraints, with the strong phases
$\delta_{\mathrm{c}}$ and $\delta_{\mathrm{n}}$ assumed as uniformly
distributed over $[-180^{\circ},180^{\circ}]$.

Taking the decay $B^0 \rightarrow K^{+} \pi^{-}$ as an example,
Fig.~\ref{fig:gammadeltaA-020} shows the constraint determined by a precise
measurement of $a_{\mathrm{CP}}(K^+\pi^-) = -0.20$ (examples with different
central values are qualitatively very similar). As can be seen, a precise
enough CP violation measurement would exclude a large portion of the
$\delta_{\mathrm{n}},\gamma$ plane, being especially effective as a
constraint on the strong phase; while the positive sign of
$\delta_{\mathrm{n}}$ is selected in the case considered, the measurement of
opposite sign, $a_{\mathrm{CP}}(K^+\pi^-) = +0.20$, represented by the same
plot after reflection with respect to the $\delta_{\mathrm{n}} = 0$ axis,
would point to $\delta_{\mathrm{n}} < 0$.

As regards the determination of $\gamma$, the main implication of a
measurement of CP violation with large enough central value would be the
possibility of excluding two (almost symmetrical) regions around $\gamma=0$
and $\gamma=180^{\circ}$. Fig.~\ref{fig:gammaA_max} shows the intervals
allowed for $|\gamma-90^{\circ}|$ at the 68\% and 95\% C.L., plotted as a
function of the measured value of the CP asymmetry $a_{\mathrm{CP}}(K^+\pi^-
)$; only the absolute value of the asymmetry is considered here, in view of
the fact that the constraint on $\gamma$ is not sensitive to the sign of the
asymmetry when the strong phase is assumed as completely indeterminate.

Measurements of the CP asymmetry in $B^+ \rightarrow K^+ \pi^0$ determine
almost the same constraints on $\gamma$ as those plotted in
Fig.~\ref{fig:gammaA_max} for $B^0 \rightarrow K^{+} \pi^{-}$, inasmuch
as the only difference in the assumed ranges for the input parameters is the
one between $r_{\mathrm{c}}$ and $r_{\mathrm{n}}$ [see Eqs.
(\ref{eq:param2}, \ref{eq:param4}, \ref{parameters})]. The CP asymmetries in
the decays $B^+ \rightarrow K^0 \pi^+$ and $B^0 \rightarrow K^{0} \pi^{0}$,
whose amplitudes depend on $\gamma$ only through the `corrective' terms
$\rho_{\mathrm{c}} e^{i (\theta_{\mathrm{c}}+\gamma)}$ and
$\rho_{\mathrm{n}} e^{i (\theta_{\mathrm{n}}+\gamma)}$ respectively [Eqs.
(\ref{eq:param1}, \ref{eq:param3})], have a minor role as constraints on
$\gamma$, but may provide a confirmation of the smallness of the
rescattering effects by placing upper limits on $\rho_{\mathrm{c}}$ and
$\rho_{\mathrm{n}}$.

\bigskip

The results of a global fit of the present $B \rightarrow K\pi$ data,
combining the CP averaged observables $R_{\mathrm{c}}$ and $R_{\mathrm{n}}$
and the direct CP asymmetries, are shown in
Fig.~\ref{fig:gammadelta_R+A_today}.

\subsection{\boldmath Constraints on the $\bar{\rho},\bar{\eta}$ plane}

\begin{figure}[htb]
\centering
\includegraphics[width=0.4\textwidth]{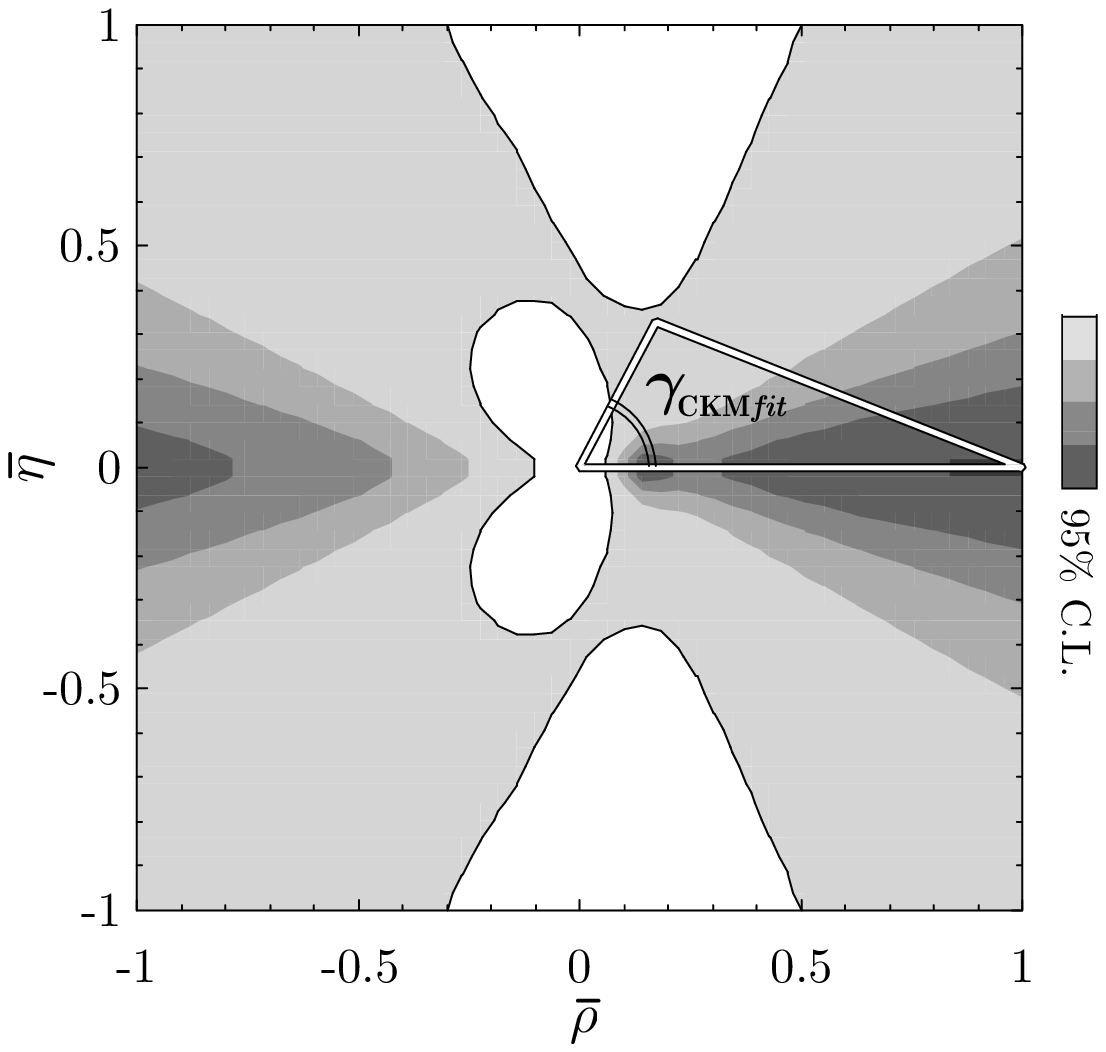}
\caption{Constraints on the vertex $\bar{\rho},\bar{\eta}$ of the UT (95\%
probability regions) determined by the combination of the current
experimental limits on the ratios of branching ratios and on the direct CP
asymmetries. A typical UT configuration favoured by the current
$|V_{ub}/V_{cb}|$, $\Delta m_{B_d}$ and $|\epsilon_{K}|$ constraints (see
Ref.~\cite{CKM}) is shown for comparison.}
\label{fig:rhoeta_today}
\end{figure}

\begin{figure}[htb]
\centering
\includegraphics[width=\textwidth]{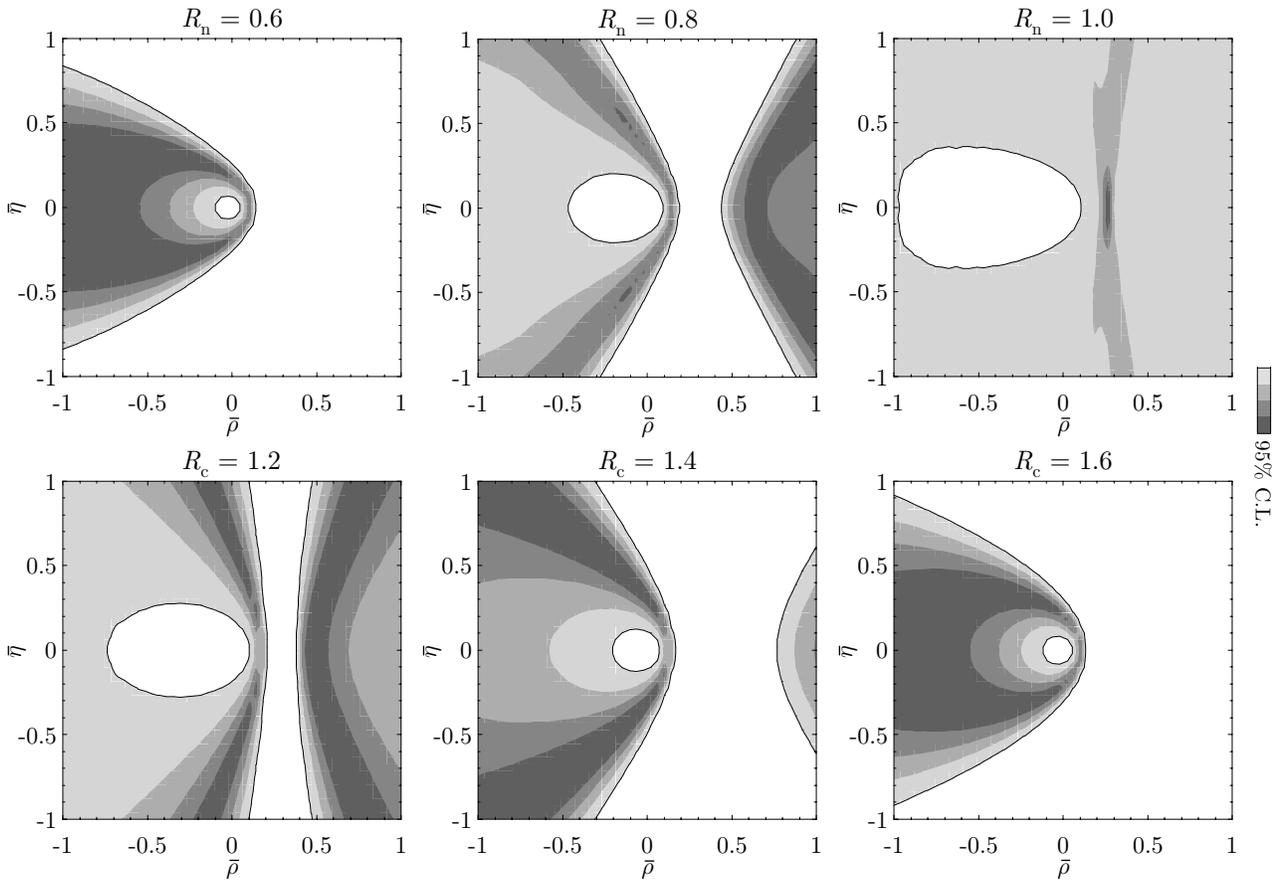}
\caption{Constraints on the vertex $\bar{\rho},\bar{\eta}$ of the UT
(95\% probability regions), as determined by precise measurements of
$R_{\mathrm{n}}$ and $R_{\mathrm{c}}$ with different central values.}
\label{fig:rhoeta_max}
\end{figure}

Measurements of $R_{\mathrm{c}}$ and $R_{\mathrm{n}}$ and of the CP
asymmetries can be included in a combined analysis of constraints on the
vertex of the UT. Instead of using the experimental determination of
$|V_{ub}/V_{cb}|$ to fix the value of the electroweak penguin parameter $q
e^{i \omega}$ [Eq.~(\ref{eq:qEW})], one can rewrite Eq.~(\ref{eq:qEW}) as
\begin{equation}
q e^{i \omega} = \frac{0.057}{\frac{\lambda}{1-
\lambda^{2}/2}\sqrt{\bar{\rho}^{2}+ \bar{\eta}^{2}}}
\end{equation}
[besides, $\gamma = \arctan(\bar{\eta}/\bar{\rho})$] and fit the available
experimental information using the variables $\bar{\rho}$ and $\bar{\eta}$.
The constraints determined by the present measurements in the
$\bar{\rho},\bar{\eta}$ plane are plotted in Fig.~\ref{fig:rhoeta_today}.
The allowed region is compatible with the results of global fits of UT
constraints \cite{CKM}, although smaller values of the CP violation
parameter $\bar{\eta}$ are favoured by the $B \rightarrow K\pi$ data.
Figure~\ref{fig:rhoeta_max} illustrates the possible effect of precise
measurements of $R_{\mathrm{c}}$ and $R_{\mathrm{n}}$.

\clearpage

\section{\boldmath Predictions for $R_{\mathrm{n}}$ and $R_{\mathrm{c}}$}
\label{ratios}

\begin{figure}[htb]
\centering
\includegraphics[width=\textwidth]{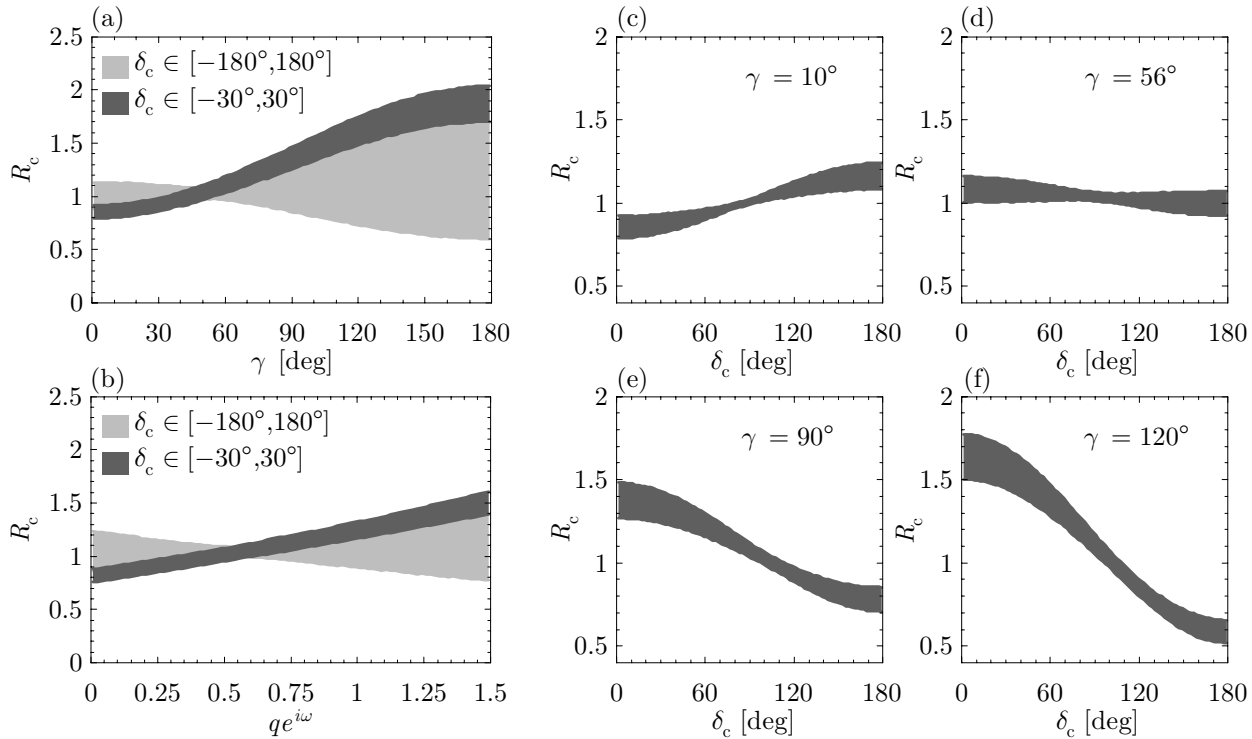}
\caption{$R_{\mathrm{c}}$ as a function of $\gamma$ (a), of the electroweak
penguin parameter $q e^{i \omega}$ (b) and of the strong phase
$\delta_{\mathrm{c}}$ for different values of $\gamma$ (c, d, e, f). The
intervals plotted for $R_{\mathrm{c}}$ are $\pm 1 \sigma$ ranges.}
\label{fig:Rc(parameters)}
\end{figure}

As is illustrated by the examples shown in the previous Section,
it is not possible to derive effective constraints on $\gamma$
\emph{in the first quadrant} from precise measurements of
$R_{\mathrm{c}}$ and $R_{\mathrm{n}}$, unless the strong phases
$\delta_{\mathrm{c}}$ and $\delta_{\mathrm{n}}$ are known to
assume certain fixed values. This limitation is the consequence of
a destructive interference between tree and electroweak penguin
amplitudes which occurs when $\gamma < 90^{\circ}$ and is maximal
for the specific value assumed by $q e^{i \omega}$ in the SM
[Eq.~(\ref{eq:qEW})]. This accidental compensation results in a
reduction of the sensitivity of $R_{\mathrm{c}}$ and
$R_{\mathrm{n}}$ to the parameters $\gamma$ and $q e^{i \omega}$
and to the strong phases [see Eqs.(\ref{eq:param1},
\ref{eq:param2}, \ref{eq:param3}, \ref{eq:param4})]. This feature
is illustrated in Fig.~\ref{fig:Rc(parameters)}, which shows the
dependence of $R_{\mathrm{c}}$ on the variables $\gamma$, $q e^{i
\omega}$ and $\delta_{\mathrm{c}}$; for each scanned value, the
remaining parameters are varied as indicated in
Eq.~(\ref{parameters}); only the real values of $q e^{i \omega}$
are considered for simplicity. As can be seen, the spread of
values of the ratio $R_{\mathrm{c}}$ is reduced considerably just
next to the SM value of $q e^{i \omega} \simeq 0.65$
[Fig.~\ref{fig:Rc(parameters)}(b)] and for $\gamma$ in the first
quadrant [(Fig.~\ref{fig:Rc(parameters)}(a)]. The wide variability
outside these regions is mainly due to the assumed indeterminacy
of the strong phase, as becomes evident when the value of
$\delta_{\mathrm{c}}$ is constrained, for example, into the range
$[-30^{\circ},30^{\circ}]$ [darker plots in the
Figures~\ref{fig:Rc(parameters)}(a) and
\ref{fig:Rc(parameters)}(b)]. As a further example, the behaviour
of the function $R_{\mathrm{c}}(\delta_{\mathrm{c}})$ plotted in
Fig.~\ref{fig:Rc(parameters)}(c-f) shows that the value of the SM
prediction for $\gamma$ ($\sim 56^{\circ}$) implies a minimum
sensitivity to the strong phase with respect to lower or,
especially, higher values of $\gamma$.

\begin{figure}[htb]
\centering
\includegraphics[width=\textwidth]{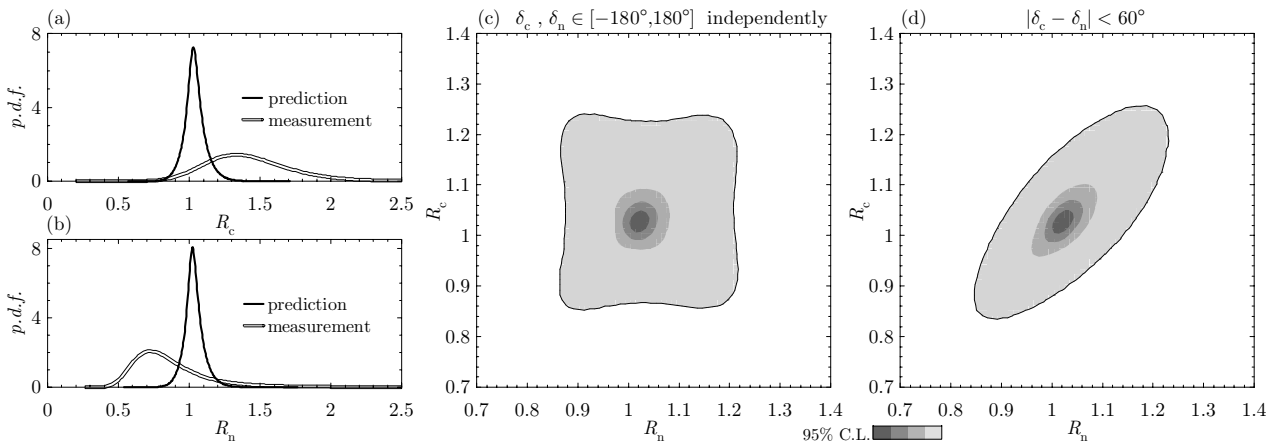}
\caption{P.d.f.'s for $R_{\mathrm{c}}$ and $R_{\mathrm{n}}$ (a, b) and
region allowed in the $R_{\mathrm{n}},R_{\mathrm{c}}$ plane for independent
(c) and correlated (d) values of the strong phases.}
\label{fig:RcRn}
\end{figure}

The effective compensation between tree and electroweak penguin
amplitudes when $\gamma$ is in the first quadrant reduces, on the
one hand, the possibility of constraining the angle $\gamma$ with
precise measurements; from a different point of view, it implies
that for $\gamma < 90^{\circ}$ the four $B \rightarrow K\pi$ decay
amplitudes are actually dominated by their common QCD penguin
components and that, consequently, values of the ratios
$R_{\mathrm{c}}$ and $R_{\mathrm{n}}$ close to~$1$ are strongly
favoured in the SM. Probability distributions for $R_{\mathrm{c}}$
and $R_{\mathrm{n}}$ obtained by assuming the SM determination for
$\gamma$ [Eq.~(\ref{eq:gamma})] are plotted in
Fig.~\ref{fig:RcRn}(a,b). They have been calculated by varying the
input parameters according to Eq.~(\ref{parameters}). The 68\%
C.L. ranges derived from these distributions are
\begin{equation}
R_{\mathrm{c}} = 1.03^{+0.07}_{-0.06} \qquad R_{\mathrm{n}} =
1.02^{+0.07}_{-0.05}, \label{results}
\end{equation}
while at the 95\% C.L. both quantities are included between $0.9$
and $1.2$. As can be seen from the $R_{\mathrm{c}} \times
R_{\mathrm{n}}$ plot shown in Fig.~\ref{fig:RcRn}(c), there is no
appreciable correlation between $R_{\mathrm{c}}$ and
$R_{\mathrm{n}}$, the ratio $R_{\mathrm{c}}/R_{\mathrm{n}}$ being
determined as
\begin{eqnarray}
& R_{\mathrm{c}}/R_{\mathrm{n}} = 1.00^{+0.08}_{-0.07}, \nonumber \\*
& 0.8 < R_{\mathrm{c}}/R_{\mathrm{n}} < 1.2 \mathrm{\ \ at\ the\ } 95\% \
\mathrm{C.L.}.
\end{eqnarray}
However, by expressing quantitatively the expectation that the strong phases
$\delta_{\mathrm{c}}$ and $\delta_{\mathrm{n}}$ should have comparable
values, the double ratio $R_{\mathrm{c}}/R_{\mathrm{n}}$ would become a very
well determined quantity: for example, the hypothetical condition
$|\delta_{\mathrm{c}} - \delta_{\mathrm{n}}| < 60^{\circ}$ leads to a
determination twice as precise:
\begin{eqnarray}
& (R_{\mathrm{c}}/R_{\mathrm{n}})_{_{|\delta_{\mathrm{c}} -
\delta_{\mathrm{n}}| < 60^{\circ}}} = 1.00^{+0.05}_{-0.04}, \nonumber \\*
& 0.9 < (R_{\mathrm{c}}/R_{\mathrm{n}})_{_{|\delta_{\mathrm{c}} -
\delta_{\mathrm{n}}| < 60^{\circ}}} < 1.1 \mathrm{\ \ at\ the\ } 95\% \
\mathrm{C.L.}.
\end{eqnarray}
The correlation between $R_{\mathrm{c}}$ and $R_{\mathrm{n}}$
introduced by this assumption is shown in Fig.~\ref{fig:RcRn}(d).

The present experimental values [Eqs.~(\ref{eq:Rc}, \ref{eq:Rn}), with
$R_{\mathrm{c}}/R_{\mathrm{n}} = 1.6^{+0.6}_{-0.5}$] are compatible with the
predictions obtained. Clearly, precise measurements are needed for a
meaningful comparison with the expected values. It has to be pointed out
that the SM expectation for $\gamma$ is not an essential ingredient of these
predictions; a simple upper limit is sufficient to obtain quite precise
values: with the only assumption that $\gamma < 90^{\circ}$, the results
\begin{eqnarray}
& R_{\mathrm{c}}|_{\gamma<90^{\circ}} = 1.02 \pm 0.10, \nonumber \\
& R_{\mathrm{n}}|_{\gamma<90^{\circ}} = 1.01^{+0.10}_{-0.09},
\end{eqnarray}
are obtained. The role of the uncertainty assumed in the present analysis to
account for possible rescattering effects ($\rho_{\mathrm{c}}$ and
$\rho_{\mathrm{n}}$ ranging from 0 up to 20\%) is also marginal in these
results, which remain essentially unchanged when $\rho_{\mathrm{c}}$ and
$\rho_{\mathrm{n}}$ are set equal to zero assuming such effects to be absent
($R_{\mathrm{c}}|_{\rho_{\mathrm{c}} = 0} = 1.03 \pm 0.06$,
$R_{\mathrm{n}}|_{\rho_{\mathrm{n}} = 0} = 1.02^{+0.06}_{-0.05}$) or when
they are fixed to the maximum value of the assumed range
($R_{\mathrm{c}}|_{\rho_{\mathrm{c}} = 0.2} = 1.03 \pm 0.07$,
$R_{\mathrm{n}}|_{\rho_{\mathrm{n}} = 0.2} = 1.02^{+0.07}_{-0.06}$). On the
contrary, as can be seen from the $R(q)$ plot in
Fig.~\ref{fig:Rc(parameters)}(b), values of $R_{\mathrm{c}}$ and
$R_{\mathrm{n}}$ inconsistent with the predictions in Eq.~(\ref{results})
may reflect large deviations from the assumed SM value of the EW parameter
$q e^{i \omega}$ [Eq.~(\ref{eq:qEW})]. To illustrate with numerical
examples, a measurement of $R_{\mathrm{c}}$ as large as $1.4$ could be
accounted for by $q e^{i \omega} = 1.2$ or, equivalently, by $q e^{i \omega}
= 0.65 e^{i 60^{\circ}}$, both leading to a determination of
$R_{\mathrm{c}}$ included between $0.7$ and $1.4$ at the 95\% C.L..
Therefore, precise measurements conflicting with the expectation
$R_{\mathrm{c}} \simeq R_{\mathrm{n}} \simeq 1$ may be the sign of large
SU(3)-breaking effects or new physics contributions to the electroweak
penguin component of the decay amplitudes.

\section{Conclusions}

Experimental constraints on the weak ($\gamma$) and strong phases of the $B
\rightarrow K\pi$ decay amplitudes have been studied in the model
independent context of the flavour-SU(3) approach. The measured rates and CP
asymmetries have been submitted to a global fit using the Bayesian method.
Possible scenarios describing the impact of precise measurements have been
reviewed in a wide range of hypothetical cases. Present situation and
prospects are summarized in the following remarks:

\begin{itemize}

\item The precision of the CP averaged data has to be increased by about one
order of magnitude in order to provide significant information on $\gamma$.
On the other hand, the first experimental limits for the direct CP
asymmetries exclude the range of values $69^{\circ} < \gamma < 111^{\circ}$
at the 95\% C.L..

\item Even within the context of minimal theoretical assumptions which
characterizes the SU(3) approach, the CP averaged observables related to $B
\rightarrow K\pi$ decays can offer interesting prospects in the search for
possible indications of new physics. Measurements of $R_{\mathrm{c}}$ and
$R_{\mathrm{n}}$ not consistent with the range $0.8 - 1.2$ would in fact
exclude the values of $\gamma$ in the first quadrant, at variance with the
UT constraints derived from the $B-\bar{B}$ oscillation parameters $\Delta
m_s$ and $\Delta m_d$. At the same time, precise measurements confirming the
currently preferred values of $R_{\mathrm{c}}>1$ and $R_{\mathrm{n}}<1$ (or
vice versa), would point to values of the strong phases
$\delta_{\mathrm{c}}$ and $\delta_{\mathrm{n}}$ belonging to two different
quadrants, in conflict with the theoretical expectation $\delta_{\mathrm{c}}
\simeq \delta_{\mathrm{n}}$.

\item On the other hand, measurements of $R_{\mathrm{c}}$ and
$R_{\mathrm{n}}$ in the range $0.8 - 1.2$, though consistent with a value of
$\gamma$ in the first quadrant, would not lead to an effective improvement
of the UT determination.

\item The strong phases $\delta_{\mathrm{c}}$ and $\delta_{\mathrm{n}}$
represent a crucial theoretical input to the analysis of the constraints on
$\gamma$; with such additional information provided by direct calculations,
a determination of $\gamma$ with $\Delta\gamma \simeq 10^{\circ}$
uncertainty becomes possible even in the least favourable case of
measurements of $R_{\mathrm{c}}$ and $R_{\mathrm{n}}$ consistent with 1.

\item On the contrary, the constraints on $\gamma$ obtainable from
$R_{\mathrm{c}}$ and $R_{\mathrm{n}}$ are almost independent of
the actual importance of rescattering effects, in so far as these
are accounted for by values of $\rho_{\mathrm{c}}$ and
$\rho_{\mathrm{n}}$ up to 0.2.

\end{itemize}

As an especially interesting result of the model independent
phenomenological analysis that has been performed, well determined SM
reference values are obtained for $R_{\mathrm{c}}$ and $R_{\mathrm{n}}$ when
$\gamma$ is fixed to its SM expectation:
\begin{eqnarray}
R_{\mathrm{c}} = 1.03^{+0.07}_{-0.06}, &
R_{\mathrm{n}} = 1.02^{+0.07}_{-0.05}. \nonumber
\end{eqnarray}
These predictions rely mainly on the SU(3) estimates of the ratios
of tree-to-QCD-penguin and of electroweak-penguin-to-tree
amplitudes, being especially sensitive to the electroweak penguin
component. They are on the other hand almost unaffected by the
possible contribution of rescattering processes and only weakly
dependent on the value assumed for $\gamma$ in the first quadrant.
The expected improvement in the experimental precision will
therefore offer the possibility of performing an interesting
experimental test of SU(3) flavour-symmetry in the decays of $B$
mesons. At the same time, precise measurements definitely
contradicting the expectation $R_{\mathrm{c}} \simeq
R_{\mathrm{n}} \simeq 1$ should lead to the investigation of
possible new physics effects in the electroweak penguin sector.

\subsection*{Acknowledgments}

We are indebted to Robert Fleischer for useful discussions on the
subject and most constructive comments on this work.

\end{document}